\documentclass[twocolumn,showpacs,preprintnumbers,amsmath,amssymb]{revtex4}
\usepackage{epsfig} 
\usepackage{dcolumn} 
\usepackage{bm}
\usepackage{pstcol,pst-grad}
\usepackage{pspicture}

\begin{document}
\title{Hysteresis and  avalanches  in the $T=0$ random-field Ising model with 2-spin-flip dynamics.}
\author{Eduard Vives}
\email{eduard@ecm.ub.es}
\affiliation{  Dept.   d'Estructura i  Constituents  de la  Mat\`eria,
Universitat  de Barcelona  \\ Diagonal  647, Facultat  de F\'{\i}sica,
08028 Barcelona, Catalonia}
\author{Martin Luc Rosinberg }
\author{Gilles Tarjus}
\affiliation{  Laboratoire  de   Physique  Th\'eorique  des  Liquides,
Universit\'e  Pierre et  Marie Curie,  4 Place  Jussieu,  75252 Paris,
France}
\date{\today}

\begin{abstract}
We study the non-equilibrium behavior of the three-dimensional Gaussian random-field Ising model at $T=0$ in the presence of a uniform external field using a $2$-spin-flip dynamics. The  deterministic, history-dependent evolution of the system is compared with the one obtained with the standard $1$-spin-flip dynamics used in previous studies of the model.  The change in the dynamics yields a significant suppression of coercivity, but the distribution of avalanches (in number and size) stays remarkably similar, except for the  largest ones that are responsible for the jump in the saturation magnetization curve at low disorder in the thermodynamic limit. By performing a finite-size scaling study, we find strong evidence that the change in the dynamics does not modify the universality class of the disorder-induced phase transition.
\end{abstract}


\maketitle

\section{Introduction}
\label{Intro}

The non-equilibrium random field Ising model (RFIM) was introduced by Sethna {\it et al.}\cite{SDKKRS1993} as a model for the Barkhausen effect in ferromagnets and more generally 
 as a prototype for many experimental systems
that show hysteretic and jerky behavior when driven by an external force. 
Because of the presence of disorder, these systems have a ``complex free energy landscape"
with a multitude of local minima (or metastable states) separated by sizeable
barriers, which makes thermally activated processes essentially irrelevant at low enough temperature (the lifetime of  metastable states may then be considered as infinite). As a consequence, these  systems  remain far from equilibrium on the experimental time scales (even when the driving rate goes to zero) and their response to the external force is made of a series of jumps (avalanches) between neighboring metastable states. This type of behavior is very well modeled by the ferromagnetic RFIM with a zero-temperature single-spin-flip  dynamics in which a spin flips only if this lowers its energy. The local character of the energy minimization is then at the origin of irreversibility. With this dynamics,  the RFIM  satisfies the property of return-point memory (or ``wiping out'' effect) which is a feature observed in several experimental systems with good approximation. Moreover, in  dimension $d\ge 3$, the model is known to exhibit an out-of-equilibrium  phase transition between a strong-disorder regime where the magnetization hysteresis loop is smooth on the macroscopic scale and a weak-disorder one where it has  a discontinuous jump. Such
a transition has been observed in thin Co/CoO films\cite{BIJPB2000} and
Cu-Al-Mn alloys\cite{M2003}, and it has been
 recently suggested\cite{DKRT2003} that it may also be associated to the 
change in the adsorption behavior of $^4$He in dilute silica aerogels\cite{TYC1999}.  The two regimes, strong and weak disorder, are separated by a critical point characterized by universal exponents and scaling laws which have been extensively studied by analytical and numerical methods\cite{DS1996,PDS1999}.  In particular, much effort has been  recently devoted  to analyze the number of avalanches, their size, and their geometrical properties above, below, and at criticality\cite{PRV2003,PRV2004}. These results provide a comprehensive, though rather complex scenario for the phase diagram of the  non-equilibrium RFIM with a metastable dynamics in the thermodynamic limit. A recent discussion of the relevance of the model to the description of the Barkhausen effect in real magnets can be found in Refs.\cite{DZ2004,SDP2004}. 

An issue that so far  has not been studied is the  robustness of this theoretical  description with respect to a change in the dynamics (in the literature on the RFIM, it is implicitely taken for granted that this should not matter). The  single-spin-flip dynamics, however, is not the unique (and may be not the best) way of simulating hysteretic dynamical processes in actual systems. It is clear for instance that the hysteresis loop will shrink if the dynamics  allows for a better equilibration of the system by employing multiple-spin flips.  Then, what will be the avalanche properties ? Will there still be a phase transition ? If so, will the critical behavior be the same as with the single-spin-flip dynamics ? There is in fact the intriguing possibility, supported by numerical simulations  and analytical arguments\cite{DS1996,PRV2004,CADMZ2004}, that the non-equilibrium and equilibrium transitions of the $T=0$ RFIM belong to the same universality class,  even if criticality occurs in zero external field at equilibrium  and at a non-zero coercive field in the irreversible evolution\cite{note1}. Since the ground state is stable with respect to the flip  of an arbitrary (finite) number of spins, this may indicate that the disorder-induced transition  has a universal character at criticality which does not depend on the specific choice of the dynamics\cite{PRV2004}. 

In  order to shed some light on this issue and check the robustness of the transition,  we study here the non-equilibrium $T=0$ RFIM with a $2$-spin-flip dynamics. We compare the results with those obtained with the standard $1$-spin-flip  dynamics, in particular those concerning the number and size  of the avalanches. We first show in section II that one can indeed define a $2$-spin-flip algorithm that yields  a  deterministic  evolution of the system with the external field. In particular, the dynamics satisfies  the  ``abelian''  property which guarantees that the same $2$-spin-flip stable configuration is attained, whatever the order in which the unstable spins are relaxed during an avalanche. It also has the  property of return-point memory. In section III we present the results of  our numerical simulations on 3-d lattices and compare them to the behavior of the same system with a $1$-spin-flip  dynamics. The hysteresis loops are significantly reduced when allowing $2$-spin flips, but the main features, including the presence of a disorder-induced transition with an associated critical point, are not significantly altered. We then perform in section IV a finite-size scaling analysis, which allows for a determination of the critical properties. We find that, within statistical uncertainty, the exponents and scaling functions are identical to those obtained with the standard $1$-spin-flip dynamics. The main conclusions of the study are reported in section V.

\section{Model and dynamics}
\label{Model}

The model  is defined on a cubic lattice of linear size $L$
with periodic boundary conditions.  On each  site ($i=1, \dots,
N=L^3$)  there is  an Ising  spin  variable ($S_i=\pm  1$).  The
Hamiltonian is
\begin{equation}
\label{Hamiltonian}
{\cal H}=-J  \sum_{<i,j>}  S_i S_j -\sum_{i} h_i S_i -H \sum_{i} S_i\; \; ,
\end{equation}
where the first sum extends over all distinct pairs of nearest-neighbor (n.n), $H$
is the  external applied field, and $h_i$ are quenched  random fields
drawn independently from  a  Gaussian distribution with zero mean and standard deviation $\sigma$.
We  are interested  in studying the  sequence of  states along
irreversible paths  at $T=0$  when the system  is driven by the external
field $H$ (in the adiabatic limit corresponding to a vanishingly small rate of change of the external field). For this purpose, the Hamiltonian must be supplemented by some
dynamical rules.

\begin{figure}
\begin{center}
\begin{pspicture}(-0.5,-0.5)(8,8)
\psset{xunit=0.7,yunit=0.7}
\pssetlength{\unitlength}{7mm}
\psline(1,9.5)(9,9.5)   
\psline(5,9.7)(5,9.3)   
\psline{->}(7,10)  (7,10.5)
\psline{<-}(2,10)  (2,10.5) 
\put(4.9,9.9){$0$}
\put(9,9.7){$f_i$} 
\put(0.2,10){\Large(a)}
\put(0.2,7.5){\Large(b)}      
\psline(1,4)(9,4)      
\psline(5,0)(5,8)
\psline(4,5)(6,3)     
\psline(6,0)(6,3)(9,3)    
\psline(1,5)(4,5)(4,8)
\psline{->}(7,6)(7,6.5)                     
\psline{->}(7.5,6)(7.5,6.5)
\psline{<-}(2,6)(2,6.5)                     
\psline{->}(2.5,6)(2.5,6.5)
\psline{<-}(2,1)(2,1.5)                     
\psline{<-}(2.5,1)(2.5,1.5)
\psline{->}(7,1)(7,1.5)                     
\psline{<-}(7.5,1)(7.5,1.5)
\put(5.2,8){$f_j'$}        
\put(9,4.2){$f_i'$}        
\put(6.1,4.2){$+J$}
\put(5.1,5.2){$+J$}
\put(3.1,4.2){$-J$}
\put(4.25,2.6){$-J$}
\put(4.7,3.6){$0$}
                  
\psline[linestyle=dotted](6,5)(6,3)
\psline[linestyle=dotted](4,5)(6,5)
\psline[linestyle=dotted](4,5)(4,3)
\psline[linestyle=dotted](4,3)(6,3)
\end{pspicture}
\end{center}
\caption{\label{fig1}  Stability  diagrams   showing  the  state  with
minimum local energy  according to  the values of  the fields created by the
neighborhood (defined in the text). (a) corresponds to a  single  spin $i$ and (b) to  a pair of
neighboring spins $i,j$.}
\end{figure}
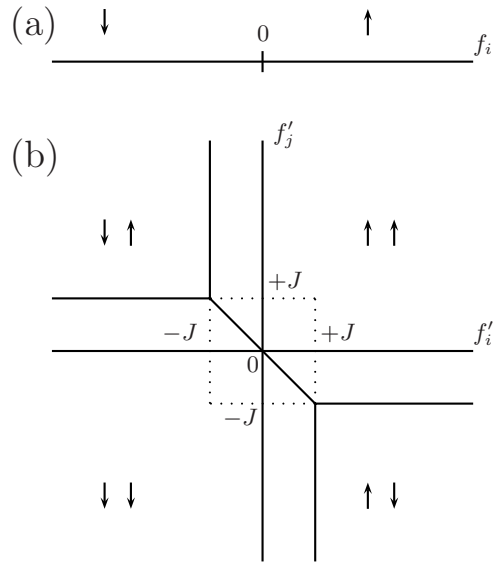
The  standard  1-spin-flip  dynamics  used  in  previous
studies consists in minimizing the energy of each spin, 
\begin{equation}
{\cal H}_i=-S_if_i
\end{equation}
where 
\begin{equation}
f_i =   J \sum_{j(i)}  S_j +h_i  + H
\end{equation}
is the net field at site $i$ (the summation in Eq.~(3) is over the $z$ n.n. of $i$).  From the above expression, it is clear that the minimization of ${\cal H}_i$ is obtained by aligning each spin with its local field, $S_i=\mbox{sign}(f_i)$, as represented schematically in Fig.~\ref{fig1}(a). This provides a stability criterium for any state with respect ot this $1$-spin-flip dynamics.  This dynamical rule may be implemented by an algorithm that  
propagates one avalanche at a time\cite{KPDRS1999}. Starting from a stable configuration, one increases (or decreases) the external field until the local field $f_i$ at some site $i$ becomes zero (this corresponds to the vanishing of the local minimum in which the system was trapped). The spin $S_i$ (which is  uniquely defined because the distribution of the random fields is continuous) is then flipped, which in turn may cause neighboring spins to become unstable and thus initiate an avalanche. The avalanche stops when a new metastable state is reached. The external field is then changed again, and so on. When the spins that become unstable during the avalanche are  sequentially reversed (e.g., by increasing $i$ from $1$ to $N$), it is  of course crucial that the final state does not depend on the sequential order.  Thanks to the ferromagnetic nature of the couplings, this is indeed the case as a result of the so-called ``no-passing" and ``abelian" properties\ of this dynamics\cite{SDKKRS1993,DSS1997}.  Moreover, the same
state  is  also reached  when all unstable spins  are flipped in parallel, which allows to measure the ``time" it takes an avalanche to occur\cite{PDS1999}.
 
\begin{figure}[ht]
\begin{center}
\epsfig{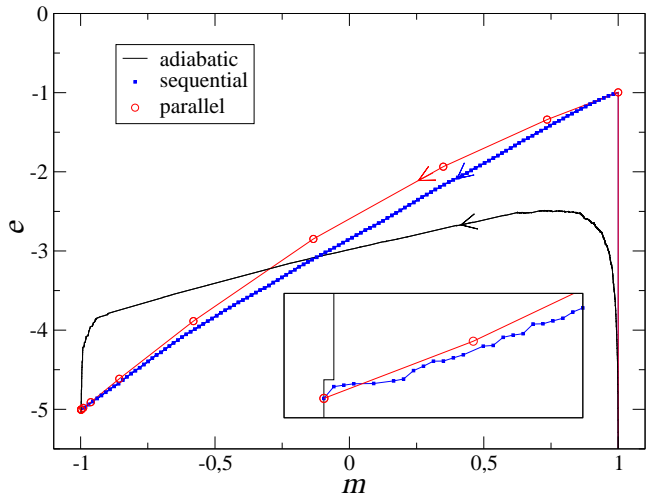}
\end{center}
\caption{\label{fig2} Evolution  of the enthalpy  and the 
magnetization (per spin) for a system  of size  $L=30$ and $\sigma=2.5$  as the external field is changed from  $H=+\infty$ to
$H=-2$.   The trajectories  corresponding to  sequential  and parallel
updating algorithms are also compared with the adiabatic one. Points indicate the intermediate unstable states. The inset shows a closeup of the region around the endpoint. Notice that the end-point is the same in all cases.}
\end{figure}
Setting the rules for a $2$-spin-flip dynamics is rather straightforward.  By definition, $2$-spin-flip stable states are spin configurations whose energy (defined solely by the Hamiltonian) cannot be lowered by the flip of one or two spins (clearly, new features are only introduced when these two spins are n.n.). The (local) energy to be minimized is thus the one associated with a pair $ij$ of n.n. spins,

\begin{equation}
{\cal H}_{ij} = -S_i f_i' -S_j f_j' -J S_i S_j 
\label{stab2}
\end{equation}
where 
\begin{equation}
f'_i= J \sum_{k(i)\neq j}  S_k +h_i  + H 
\end{equation}
is  the field experienced by $S_i$  {\it without} the influence of the neighbor $S_j$ (the summation in Eq. (5) is over the n.n. of $i$ excepting $j$).  One can then think of the dynamics as made of single-spin flips and ``irreducibly cooperative" $2$-spin flips. As pictured in Fig.
\ref{fig1}(b), a single-spin flip occurs whenever the net field on $S_i$, $f_i=f'_i+JS_j$, or on $S_j$, $f_j=f'_j+JS_i$, changes sign. This corresponds to the changes $\downarrow \downarrow \leftrightarrow \uparrow \downarrow$, $\downarrow \downarrow \leftrightarrow \downarrow \uparrow$, $\uparrow \uparrow \leftrightarrow \uparrow \downarrow$, $\uparrow \uparrow \leftrightarrow \downarrow \uparrow$ in the diagram.
An irreducibly cooperative 2-spin flip involves a n.n. pair of spins with the same sign ($\uparrow \uparrow $ or $\downarrow \downarrow $) that cannot flip individually (i.e. without the simultaneous flip of the neighbor). This occurs whenever the net field on the pair of aligned spins, $f_{ij}=f'_i+f'_j$, changes sign in the region of the diagram where $-J\leq f'_i\leq J$ and $-J\leq f'_j\leq J$. These two last conditions come from the fact that the spins cannot individually flip (hence neither $f_i$ nor $f_j$ changes sign before the cooperative flip of the pair) and the condition that once a pair has flipped, none of the spins can flip back individually. 

The corresponding algorithm is a simple extension of the one described above for the $1$-spin-flip dynamics. Starting from a $2$-spin-flip stable configuration, the external field is varied until one finds a pair of spins that becomes marginally stable: the representative point of this pair in the diagram of Fig.~\ref{fig1}(b) leaves the region where it was  originally (associated with $\uparrow \uparrow$, $\uparrow \downarrow$, $\downarrow \uparrow$, or $\downarrow \downarrow$) and, depending on the border which is first attained, only one spin flips or the two spins flip simultaneously\cite{note2}. 

It is not hard to show that this new dynamics obeys the same properties as the $1$-spin-flip dynamics, in particular the crucial abelian property. This is again a consequence of the ferromagnetic nature of the interactions. One only needs
to note that the state of the system can be represented
by a set  of $zN/2$ points (corresponding to  all the distinct n.n.
pairs)  in the  diagram of Fig.~\ref{fig1}(b).   As the external field is monotonously increased (resp. decreased),  the local fields can only increase (resp. decrease), the spins can only flip up (resp. down), and the points can only move up and right  (resp. down and left) in the diagram.
By slightly modifying the arguments of Ref.~\cite{SDKKRS1993}, one then can
prove the no-passing rule, the abelian property and the  existence of return-point memory. 

\begin{figure}[ht]
\begin{center}
\epsfig{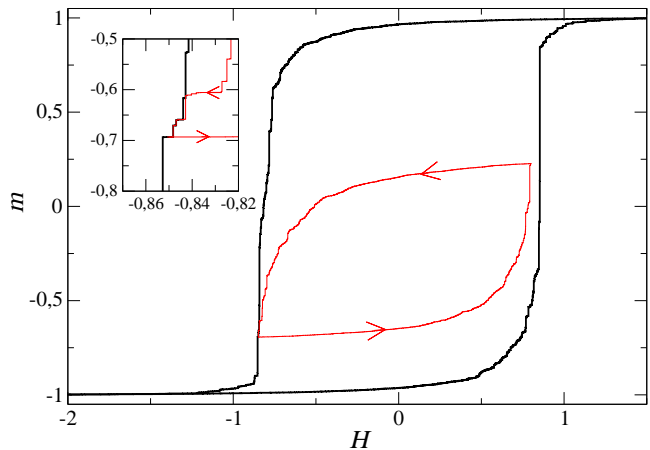}
\end{center}
\caption{\label{fig3}   Major  hysteresis   loop   and an  internal   loop
($-0.85\leq H \leq0.8$)  obtained  with the  2-spin-flip   dynamics for a system  of size $L=30$  and $\sigma=2.5$.   The inset
shows  the details around  the   point at $H=-0.85$ revealing  the property of return-point memory.}
\end{figure}
Instead of paraphrasing the demonstrations given in Ref.\cite{SDKKRS1993},  we choose here to illustrate 
these properties by a numerical example. (We  have  also performed 
numerical tests in many situations and found no violations of these properties.)
The evolution of a system
with size $L=30$ and $\sigma=2.5$ is shown in Fig.~\ref{fig2} where the energy per spin, $e={\cal H}/N$, is plotted as
a function  of the magnetization  $m=\sum S_i /N$ (strictly speaking, $e$ is the enthalpy). The  external field
$H$ is  varied from  a very  large initial value  where all spins are up to the  final value $H=-2$ (here and after, $J$ is taken as the energy unit).  Two of the curves display the  sequence of  unstable states that are 
obtained after a  sudden change of the external field  using either a sequential or a
parallel updating  algorithm. We  also show the  metastable evolution
corresponding  to  the adiabatic  driving  (with sequential  updating)
along the  hysteresis loop. In all  cases the final state  is the same. This is true even when the intermediate  states are distinct,  for instance when the
spins are chosen sequentially in  a different order. 
 
The property of return-point memory property is illustrated in Fig.~\ref{fig3},
again for a  system  with size $L=30$  and $\sigma=2.5$.  The minor loop  is obtained by reversing  the evolution of the external field first in the decreasing
branch  at $H=-0.85$ and  then at $H=+0.80$.  As  can be  seen from the inset, the  internal loop
closes before the return point,  so that the evolution 
follows that of the major loop for a small region of $H\gtrsim-0.85$.

\section{Numerical simulations}
\label{Numerical}

\begin{figure}[ht]
\begin{center}
\epsfig{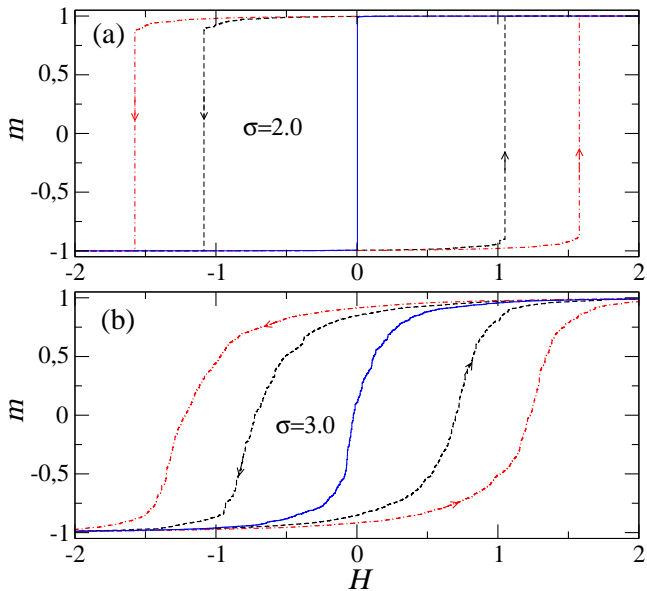}
\end{center}
\caption{\label{fig4}   Magnetization curves obtained with the $1$-spin and  $2$-spin-flip dynamics in a sample of size $L=30$  for (a)
$\sigma=2$ and (b) $\sigma=3$:  dotted-dashed lines correspond to the $1$-spin-flip dynamics and dashed lines to the $2$-spin-flip dynamics. In addition,  the ground-state magnetization is shown as continuous lines.}
\end{figure}
We now present  the results  of numerical simulations performed on 3-dimensional cubic lattices using  either the $1$-spin-flip or $2$-spin-flip dynamics with sequential updating.  Averaged quantities were obtained with statistics over $10^3$ to $10^5$ different realizations of the random field distribution and system sizes ranging from  $L=8$ up to $L=48$. As emphasized in Ref.\cite{PRV2003}, in order to describe properly avalanche properties (especially  the ``spanning'' avalanches), it is more important to perform averages  over many disorder realizations than to simulate very large system sizes.

\subsection{Hysteresis loops}

\begin{figure}[ht]
\begin{center}
\epsfig{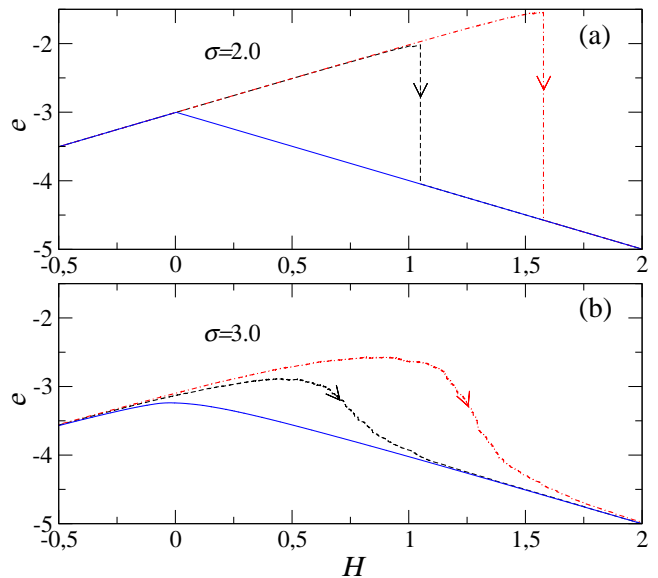}
\end{center}
\caption{\label{fig5} Enthalpy per spin as a function of $H$ corresponding to the loops in Fig.~5 (for clarity, only the ascending branches are shown). The continuous lines represent the ground-state behavior.}
\end{figure}
Fig.~\ref{fig4} shows the hysteresis loops obtained in a single sample for two different  values of the disorder $\sigma$.  For comparison, we also display the magnetization curves obtained with the $1$-spin-flip dynamics and with the algorithm of Ref.\cite{FGOV2000} which gives  the exact ground-state (equilibrium) magnetization. The corresponding behavior of the enthalpy per spin along the ascending branches of the loops is reported in Fig.~\ref{fig5} (note in passing that the ground-state enthalpy does not show any discontinuity as the external field is varied\cite{FGOV2000}).  As could be expected, the main effect of the new dynamics is to reduce the size of the hysteresis loops. Specifically,  the coercivity (i.e., the magnitude of the external field for which the magnetization is equal to zero) is decreased by more than $30\%$ when allowing  pairs of spins to flip together. Accordingly, the enthalpy difference between the ground state and the metastable states  which are visited along the loops is also reduced.  Nevertheless, the loops display the same key feature, that is a change from a discontinuous to a continuous behavior as the disorder is increased.  This suggests  that there is also an out-of-equilibrium disorder-induced phase transition under the  $2$-spin-flip dynamics, with a critical value of  $\sigma$ at which the discontinuity appears for the first time in the thermodynamic limit.

It is worth pointing out that the new dynamical rules allows the system to effectively overcome energy barriers of magnitude up to $\Delta E=2J$. Indeed, the difference between the two dynamics shows up when a pair of n.n. spins with same sign can cooperatively flip, say from  $\downarrow \downarrow$ to $\uparrow \uparrow$ when the external field is increased, whereas each of its spin cannot individually flip. This means that along $1$-spin-flip paths, the system has now been able to bypass the 
higher-energy states, either $\uparrow \downarrow$ or $\downarrow \uparrow$. By using Eq.~(4), it is easy to show that the relevant barrier height associated with this process is at most $2J$. Since cooperative flips of more than $2$ spins do not occur with the chosen dynamics and the system's trajectory otherwise go through states of decreasing energy, one concludes that $\Delta E=2J$ is the maximum barrier height that the system may overcome when passing from the $1$-spin-flip to the $2$-spin-flip dynamics.

\subsection{Avalanches}

As shown in recent studies\cite{PRV2003,PRV2004}, a good characterization of the  disorder-induced critical point can be reached by 
analyzing the number and size distribution of  the magnetization jumps (avalanches) that compose  the hysteresis loops in finite systems.  For that purpose, it is necessary to classify the avalanches in several  categories, according to their behavior as the system size $L$ is increased.  One first has to distinguish  whether or not an avalanche spans the  system from one side to the other,  in $1$, $2$ or $3$ spatial directions (indicated in the following by the index $\alpha$). For each individual avalanche, this  is a property that can be easily detected during the simulation. Avalanches are thus classified as being  non-spanning  ($\alpha=$ns), $1$d-spanning  ($\alpha=1$),  $2$d-spanning ($\alpha=2$), or $3$d-spanning  ($\alpha=3$).  

\begin{figure}[ht]
\begin{center}
\epsfig{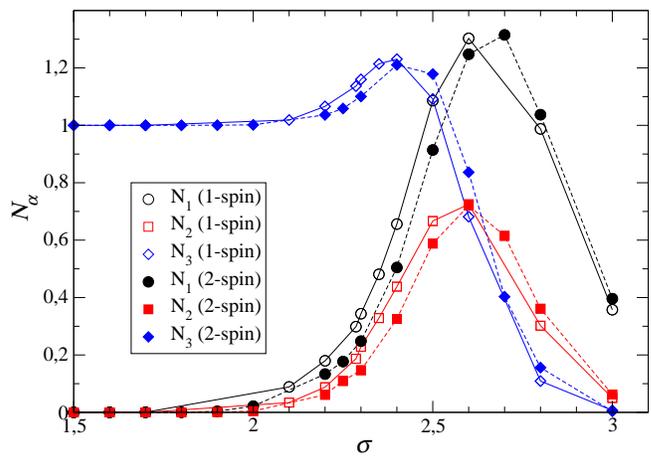}
\end{center}
\caption{\label{fig6}   Average  numbers   of   $1$d, $2$d, and $3$d-spanning avalanches as  a  function  of  $\sigma$ in  a  system  of size  $L=24$.
Continuous and dashed lines  are guides for the eye and correspond to the $1$-spin-flip and $2$-spin-flip dynamics, respectively.}
\end{figure}

Fig.~\ref{fig6} shows  the number
of $1$d, $2$d and $3$d-spanning avalanches recorded along the descending branch of the hysteresis loops as a function of  $\sigma$.   The data, averaged  over disorder,  correspond to  a  system  of size
$L=24$.   It can be seen that  the behavior of  the three quantities  is completely 
equivalent  under the two dynamics. The only  difference is a shift towards  larger values of
$\sigma$ when  the $2$-spin-flip dynamics  is used. The same  shift is also
found for  all studied system  sizes.  This is  a first indication
that   $\sigma_c^{(2)}>  \sigma_{c}^{(1)}$,  as will be confirmed 
by the  finite-size  scaling analysis presented in the next section. 
In  the case of the $1$-spin-flip  dynamics, a detailed  analysis was performed in Refs.\cite{PRV2003,PRV2004}, revealing the  scenario that occurs  in  the thermodynamic limit and that is already suggested by the data shown in Fig.~\ref{fig6}: when $L \rightarrow \infty$, $N_1(\sigma)$ and $N_2(\sigma)$ are expected to display a $\delta$-singularity at $\sigma_c$ and $N_3(\sigma)$ a step-like behavior.  It  was 
shown, moreover,  that  there are two types of  3d-spanning avalanches, {\it subcritical} and  {\it critical}, which  scale with different exponents. The former are 
responsible for  the discontinuity in the magnetization curve in the thermodynamic limit (there is only one, compact, subcritical avalanche for $\sigma< \sigma_c$) whereas the latter only exist at  $\sigma_c$ (hence the additional $\delta$-singularity at the edge of the step function whose signature is already visible in Fig.~\ref{fig6}).  In a finite system, however, all  kinds of avalanches may exist close enough
to the critical point, and it is quite difficult to discriminate subcritical from critical avalanches.  In Ref.\cite{PRV2004}, an elaborate analysis  was needed to  show that these avalanches have different
fractal dimensions at criticality. This study is impossible here because of the complexity of the $2$-spin-flip algorithm that forbids the use of large systems with good enough statistics. Therefore, in the following, we shall not distinguish between these subcritical and critical $3$d-spanning avalanches.

\begin{figure*}[ht]
\begin{center}
\epsfig{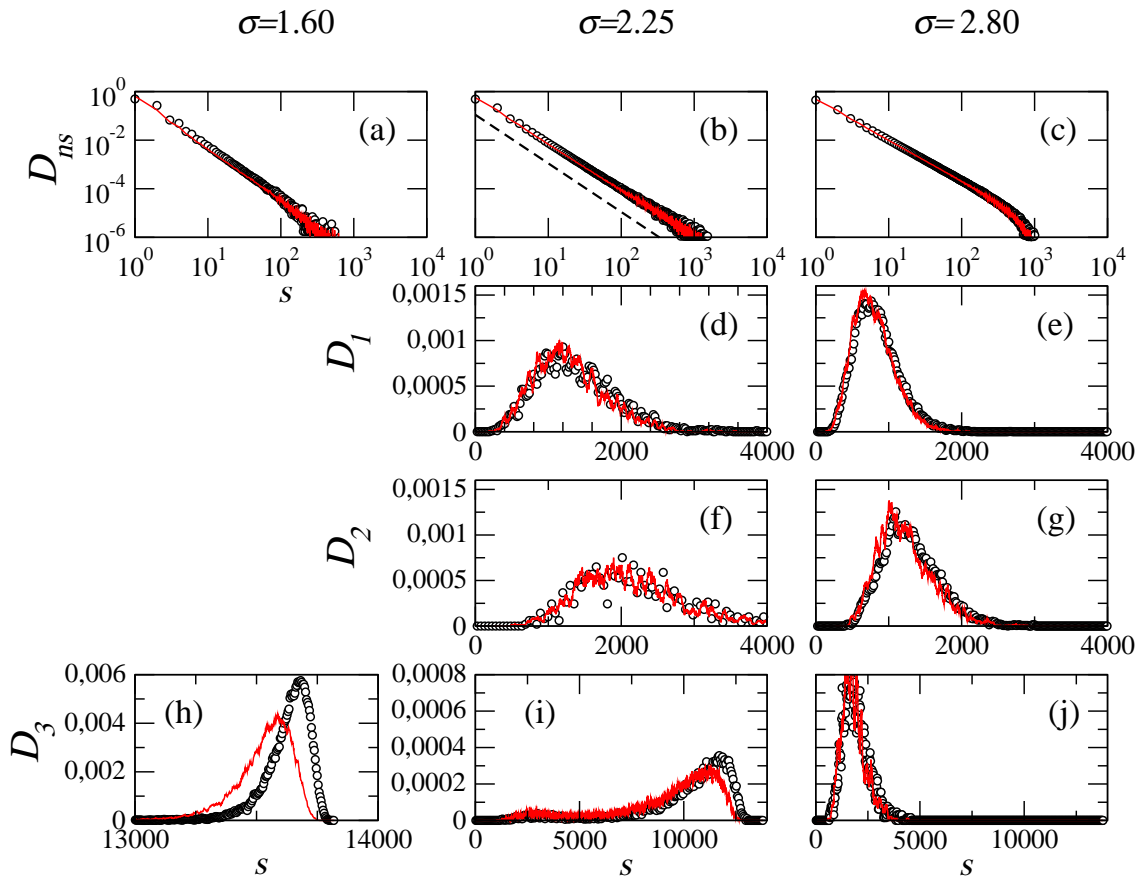}
\end{center}
\caption{\label{fig7}  Avalanche size distributions  for $\sigma=1.60$
(a,h), $2.25$ (b,d,f,i), and $2.80$ (c,e,g,j). The open symbols  and the continuous lines correspond to the $2$-spin-flip and $1$-spin-flip dynamics, respectively.   The first
row  shows the  distribution  $D_{ns}$ of  non-spanning  avalanches  and the other rows show the  distributions  $D_1, D_2$ and   $D_3$ of  the $1$d, $2$d and $3$d-spanning avalanches, respectively. All  data have been obtained in a system of  size $L=24$. In (b), the
dashed line indicates the expected power-law behavior of $D_{ns}$ at criticality with exponent $\tau_{eff}=2.0$.} 
\end{figure*}

Fig.~\ref{fig7} shows the (normalized) avalanche size distributions
$D_{\alpha}(s;\sigma,L)$  obtained along one branch of the hysteresis loop for  three
different  values  of  $\sigma$ (for clarity, the results obtained with the 
 $1$-spin-flip dynamics are represented by continuous lines). 
Surprisingly, one can see that the distribution  of  non-spanning avalanches   in Figs.~\ref{fig7}(a)  to \ref{fig7}(c) is almost unaltered by the change in the dynamics.
The only small difference (barely visible on the figure) induced by the $2$-spin-flip dynamics is 
that there  are a little less avalanches  of size
$s=1$ and a little more avalanches  of size $s=2$, but the rest of the
distribution is  almost the same.  In particular,  with both  dynamics, the expected power-law
behavior  of the distribution will be  characterized by  the same  exponent $\tau_{eff}\approx 2.0$  in the thermodynamic limit\cite{PRV2004}.

The   size  distributions  $D_1$ and $D_2$ of   the    $1$d and $2$d-spanning
avalanches shown in Figs.~\ref{fig7}(d) to  \ref{fig7}(g) also appear to be identical with the two dynamics, at least within statistical error bars. (Note that these avalanches do not exist for $\sigma=1.6$ because this value is  much lower  than $\sigma_c$ for both  dynamics.) The  only visible
differences between the two dynamics occur  in $D_3$, the  size  distribution  of   the   $3$d-spanning avalanches.  Specifically,  for $\sigma=1.60$    and $2.25 $ (i.e. below $\sigma_c$ and very close to $\sigma_c$, respectively), the  large 3d-spanning  avalanches tend to be shifted to even larger sizes. According to Refs.~\cite{PRV2003,PRV2004}, these avalanches 
are probably {\it subcritical}  spanning avalanches and their average size is thus a measure  of the  order  parameter. Therefore, this  result  is another indication that   $\sigma_c^{(2)} > \sigma_c^{(1)}$.  In contrast, for $\sigma=2.80$ (which is clearly above $\sigma_c$), 
the distribution $D_3$
is not  affected by  the dynamics (Fig.\ref{fig7}(j)): in this  case, one expects to  detect  
only  {\it critical} 3d-spanning  avalanches in a finite system.

\begin{figure*}[ht]
\begin{center}
\epsfig{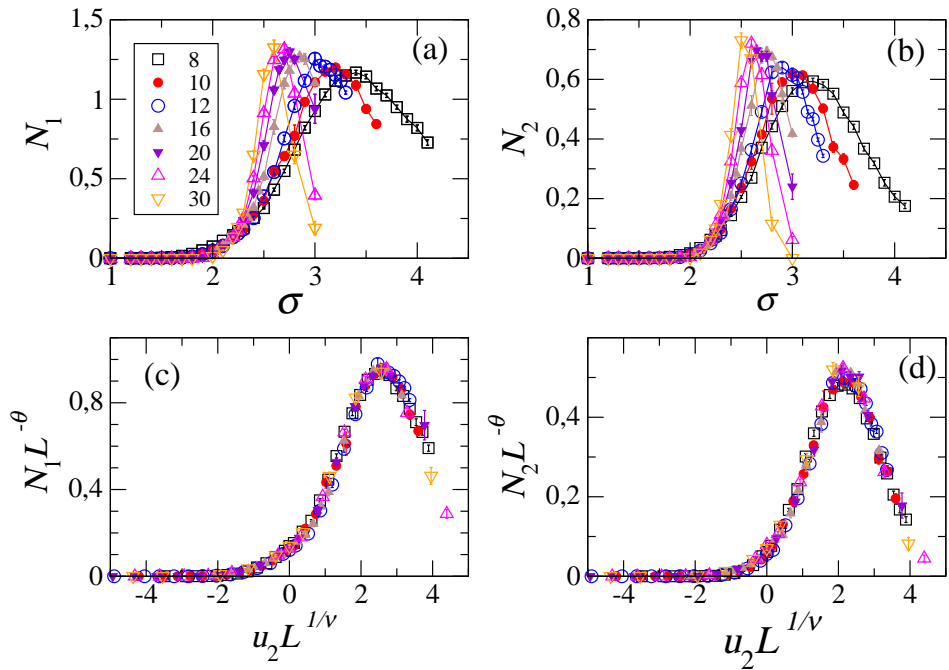}
\end{center}
\caption{\label{fig8} Number of spanning avalanches in $1$d (a) and $2$d (b) obtained with the $2$-spin-flip  dynamics for different system sizes. The corresponding scaling plots  in (c)  and  (d) have been obtained according to Eqs. (9-11) with 
$\sigma_c^{(2)}=2.25$, $A=-0.2$, $\nu=1.2$ and $\theta=0.1$.}
\end{figure*}

\section{Finite-Size Scaling analysis}

As already mentioned, a precise determination of $\sigma_c^{(2)}$  
and of some of the critical exponents  requires a detailed finite-size  scaling  analysis 
which, unfortunately, is not posssible with the  present algorithm. What can be done, however, is 
to check whether the set of exponents found with the $1$-spin-flip dynamics can also be used to scale
the new $2$-spin-flip data.  We first consider  $N_1$ and $N_2$, the numbers of  $1$d and $2$d-spanning avalanches, which are quantities that have a
simple  scaling behavior. Following previous work\cite{PRV2003,PRV2004}, we assume the forms
\begin{equation}
N_{1}(\sigma,L) = L^{\theta} {\tilde N}_1 \left ( u_2L^{1/\nu} \right )
\end{equation}
\begin{equation}
N_{2}(\sigma,L) = L^{\theta} {\tilde N}_2 \left ( u_2L^{1/\nu} \right )
\end{equation}
where  $\theta$ and $\nu$ are critical exponents, $\tilde N_1$  and $\tilde  N_2$ are scaling functions, and $u_2$ is a scaling variable that 
measures the distance to the critical point. It is defined as

\begin{equation}
u_2=\frac{\sigma  -  \sigma_c^{(2)}}{\sigma_c^{(2)}}  +  A \left  (  \frac{\sigma  -
\sigma_c^{(2)}}{\sigma_c^{(2)}} \right )^2
\end{equation}
where the parameter $A$ accounts for  a second order correction that plays 
a role when the studied systems are
not very large, as is the case here.  In Ref.\cite{PRV2003}, the best choice for the collapse of the scaling plots was obtained with $\nu=1.2$, $\theta= 0.1$, and $A=-0.2$, values that we keep here. The only free parameter is thus  $\sigma_c^{(2)}$. 

As shown in Fig.~\ref{fig8}, a very good collapse of the new data can be obtained with 
$\sigma_c^{(2)}=2.25$. Taking into account the  quality  of the  plots, there is an  uncertainty of $\pm
0.01$ on this value, but, clearly, the value $\sigma_c^{(1)}=2.21$ obtained with the $1$-spin-flip  dynamics\cite{PRV2003} can be discarded. We have also checked that this conclusion is not modified when the non-universal parameter $A$ is allowed to vary.
These scaling plots thus indicate that the $2$-spin-flip dynamics only induces a  shift of the critical value  of the  disorder but does not change the 
universality class of the transition.
\begin{figure}[ht]
\begin{center}
\epsfig{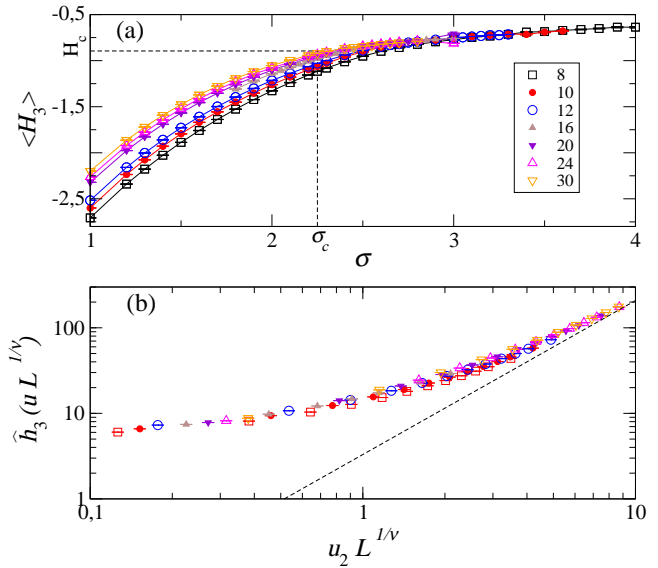}
\end{center}
\caption{\label{fig9} (a) Average field $\langle H_3(\sigma,L)  \rangle $
at which the 3d-spanning  avalanches occur as a function of $\sigma$ for different system sizes. (b) Scaling plot of the data below $\sigma_c^{(2)}$ according to Eq. \ref{H3} with $H_c^{(2)}=-0.885$, $B'=0.25$ and $\mu= 1.5$.}
\end{figure}

A second check of this result may be obtained by measuring the average
field $\langle  H_3(\sigma) \rangle$  at which the $3$d-spanning  avalanches occur.
This quantity was studied  with the   $1$-spin-flip dynamics  in
Ref.~\cite{PRV2004}, which allowed to map out the first-order
line  (corresponding to the macroscopic jump in the magnetization) in the  diagram  $H-\sigma$.   The dependence of $\langle  H_3(\sigma) \rangle$ with the system size is shown in Fig.~\ref{fig9}(a). Note that the transition line
 extends above
$\sigma_c$ in a finite system. However, the end-point, beyond which no
3d-spanning  avalanches are found (with the present sampling of disorder realizations), becomes closer and closer to the critical point $(H_c,\sigma_c)$ as $L$ is increased.

According to Ref.~\cite{PRV2004}, this set of curves should scale as
\begin{equation}
\label{H3}
\langle     H_3     \rangle    (\sigma,L)     =     H_c^{(2)}    \left     (
1-B'\frac{\sigma-\sigma_c^{(2)}}{\sigma_c^{(2)}}  \right ) -L^{-1/\mu}  {\hat h}_3
\left ( u_2L^{1/\nu} \right )
\end{equation}
where $H_c^{(2)}$ is the critical field, $B'$ a  non-universal tilting constant, $\mu$ a  critical exponent, and  $\hat h_3$ the
corresponding scaling function. Strictly
speaking,  the scaling should be done separately for the average fields $\langle     H_{3-} \rangle$ and $\langle H_{3c} \rangle$ at which 
the subcritical and critical $3$d-spanning avalanches occur\cite{PRV2004}. Indeed, the  number  of these avalanches  scales
differently with $L$. However we expect the lack of scaling to have a (small) effect only in the region $u_2 L^{1/\nu} \simeq 1$ where the two kinds of avalanches coexist\cite{PRV2004}.

In Ref.~\cite{PRV2004}, the best collapse of the $1$-spin-flip data was obtained with $\mu=1.5$, $B'=0.25$, and $H_c^{(1)}=-1.425$. Here, we keep the same values for $\mu$ and $B'$, set $\sigma_c^{(2)}=2.25$, and  consider the critical field as the only free  parameter. As shown in Fig.~\ref{fig9}(b), a very good collapse can be 
obtained with $H_c^{(2)} =-0.885$ (for clarity, we only present the collapse for $\sigma<\sigma_c^{(2)}$). Again, this  result  is   consistent  with  the  assumption that  the new dynamics does not change 
the  universality  class   of the transition.   On the other hand, the significant decrease in the critical field (in absolute value) is in line with the decrease in coercivity illustrated by Fig.~4.

\begin{figure}[ht]
\begin{center}
\epsfig{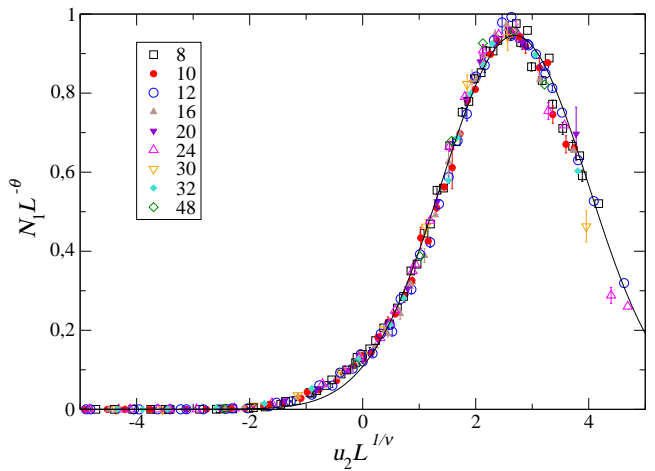}
\end{center}
\caption{\label{fig10}    Comparison   of  the    scaling   plots of the number of $1$d-spanning avalanches obtained with the $1$-spin-flip (symbols with error bars)
and $2$-spin-flip (symbols without  error bars) dynamics. The two collapses only differ by the value of the critical disorder ($\sigma_c^{(1)}=2.21$ and $\sigma_c^{(2)}=2.25$). The continuous line is the    Gaussian     fit    proposed    in Ref.~\cite{PRV2003}.}    
\end{figure}

Finally, it is interesting to study the influence of the new dynamics on the finite-size scaling functions.  
In Fig.~\ref{fig10}, we compare the scaling collapses  of the number of $1$d-spanning avalanches, $\tilde N_1(\sigma,L)$, obtained with the two dynamics (for the $2$-spin-flip dynamics, this is the same curve as in Fig.~8).
As  can   be  seen, the   agreement  between  the  two  curves is quite
remarkable. Even the deviations (around $u_2L^{1/ \nu} \approx 0)$ from  the Gaussian fit proposed in Ref.\cite{PRV2003} are the same. A similar  analysis for $\tilde N_2(\sigma,L)$ shows 
the same agreement.  This is again a  strong indication for universality, that goes beyond the equality of the critical exponents\cite{note3}.

\section{Conclusion}

We have shown in this paper that the non-equilibrium behavior of the $3$-d RFIM at zero temperature is not qualitatively altered when going from the standard $1$-spin-flip metastable dynamics to a $2$-spin-flip metastable dynamics. The coercivity (that is essentially the width of the hysteresis loop associated with the evolution of the magnetization with the driving field) is significantly reduced by allowing $2$-spin flips, but the main features of the hysteretic behavior, including the presence of a disorder-induced non-equilibrium transition and the distribution of avalanches, remain similar. By using a finite-size scaling analysis, focused on the number and size of the avalanches, we have furthermore provided strong evidence that the critical behavior (exponents and scaling functions) obtained with the $2$-spin-flip dynamics is in the same universality class as that obtained with the $1$-spin-flip dynamics. 

Changing the dynamical rules used to study the evolution of the model, as done here, helps adressing several important questions. The first one is the robustness of the hysteretic scenario provided by the $T=0$ RFIM with the standard $1$-spin-flip dynamics.  As discussed in introduction, a basic assumption underlying the theoretical description is that the system gets trapped in metastable states on the experimental time scale and can only escape when a change in the external field makes the relevant state looses its stability. However, real systems are not at $T=0$ and some partial, local equilibration, due for instance to thermally activated processes, may take place even though the system remains far from equilibrium on the experimental time scale. Introducing cooperative $2$-spin flips is a simple way to check the effect of partial equilibration processes. Our results clearly point towards the robustness of the whole theoretical picture drawn from previous studies of the model using the $T=0$ $1$-spin-flip dynamics\cite{SDKKRS1993,DS1996,PDS1999,PRV2003,PRV2004}.

A second question concerns the relation between the disorder-induced critical properties observed in the non-equilibrium behavior of the RFIM at $T=0$ and the equilibrium critical behavior associated with the paramagnetic to ferromagnetic transition. There is numerical evidence that, at least to a good approximation, critical exponents and scaling functions associated with the two kinds of criticality are the same\cite{DS1996,PRV2004,CADMZ2004}. (Additional, but less conclusive evidence is provided by exact, but mean-field-like results on the Bethe lattice\cite{CADMZ2004} and by perturbation theory near the upper critical dimension 
$d=6$\cite{DS1996}.) Our present findings suggest that a whole series of $T=0$ metastable dynamics involving $k$-spin flips lead to the same non-equilibrium criticality, with the critical disorder strength increasing with $k$ and the critical coercive field decreasing with $k$. Equilibrium behavior at $T=0$ as a function of the external field involves the system's ground state, i.e. a state stable to flips of any arbitrary finite number of spins. Based on the numerical closeness of the critical exponents and scaling functions, on the fact that the critical disorder strength 
satisfies $\sigma_c^{eq}>\sigma_c^{(2)}>\sigma_c^{(1)}$ (whereas $H_c^{eq}=0<H_c^{(2)}<H_c^{(1)}$), and on the similarity of the underlying physics at $T=0$, it is tempting to speculate that the equilibrium behavior can be obtained as the limit of a series of $k$-spin-flip metastable dynamics with increasing $k$, with the critical properties of the whole series belonging to the same universality class and governed by the same fixed point\cite{PRV2004}.

\section*{Acknowledgements}

We    acknowledge   fruitfull   discussions    with   Francisco-Jos\'e
Perez-Reche and Xavier Illa.  E.V also acknowledges the hospitality of the Laboratoire
de Physique Th\'eorique des Liquides  (UPMC, Paris) during his stay as
invited  professor in  July  2004. This  work  has received  financial
support from CICyT (Spain), project MAT2004-01291, and CIRIT (Catalonia), project 2000SGR00025. The Laboratoire de Physique Th\'eorique des Liquides is the UMR 7600 of
the CNRS.


\begin{thebibliography}{10}

\bibitem{SDKKRS1993} J. P. Sethna, K. Dahmen, S. Kartha, J. A. Krumhansl,
B. W. Roberts, and J. D. Shore, Phys. Rev. Lett. {\bf 70}, 3347
(1993). 
\bibitem{BIJPB2000} A. Berger, A. Inomata, J. S. Jiang, J. E. Pearson, and S. D. Bader, Phys. Rev. Lett. {\bf 85}, 4176
(2000).
\bibitem{M2003} J. Marcos, E. Vives, Ll. Ma\~{n}osa, M. Acet, E. Duman,
M. Morin, V. Novak, and A. Planes, Phys. Rev. B {\bf 67}, 224406
(2003).
\bibitem{DKRT2003} F. Detcheverry, E. Kierlik, M. L. Rosinberg, and
G. Tarjus, Phys. Rev. E {\bf 68}, 061504 (2003); Langmuir {\bf 20},  8006  (2004).
\bibitem{TYC1999} D. J. Tulimieri, J. Yoon, and M. H. W. Chan, Phys. Rev. Lett. {\bf 82}, 121 (1999).
\bibitem{DS1996} K. Dahmen and J. P. Sethna, Phys. Rev. B {\bf  53},
14872 (1996).
\bibitem{PDS1999} O. Perkovi\'c, K. Dahmen, and J. P. Sethna, Phys. Rev. B
{\bf 59}, 6106 (1999).
\bibitem{PRV2003} F. J. P\'erez-Reche and E. Vives, Phys. Rev. B
{\bf 67}, 134421 (2003).
\bibitem{PRV2004} F. J. P\'erez-Reche and E. Vives, cond-mat/0403754 (to appear in Phys. Rev. B {\bf 70}).
\bibitem{DZ2004} G. Durin and S. Zapperi, cond-mat/0404512, to appear in {\it The Science of Hysteresis}, edited by G. Bertotti and I. Mayergoyz, Elsevier (2004).
\bibitem{SDP2004} J. P. Sethna, K. A. Dahmen, and O. Perkovi\'c, cond-mat/0406320, to appear in {\it The Science of Hysteresis}, edited by G. Bertotti and I. Mayergoyz, Elsevier (2004).
\bibitem{CADMZ2004} F. Colaiori, M. J. Alava, G. Durin, A. Magni, and S. Zapperi, Phys. Rev. Lett. {\bf 92}, 257203 (2004); M. J. Alava, V. Basso, F. Colaiori, L. Dante, G. Durin, A. Magni, and S. Zapperi, cond-mat/0407297.
\bibitem{note1} This objection, however, may  be circumvented by considering the remanent magnetization obtained in zero external field after a demagnetization procedure as the order parameter of the transition\cite{CADMZ2004}.
\bibitem{KPDRS1999} M. C. Kuntz, O. Perkovi\'c, K. A. Dahmen, B. W. Roberts, and J. P. Sethna, Computing in Science and Engineering 1, 73 (1999).
\bibitem{DSS1997} D. Dhar, P. Shukla,  and J. P. Sethna, J. Phys. A: Math. Gen.  {\bf  30}, 5259 (1997).
\bibitem{note2} Unfortunately, because of the additional complexity introduced by the cooperative flip of n.n. pairs, one cannot use anymore the so-called {\it sorted-list} algorithm proposed in Ref.\cite{KPDRS1999}. This in turn implies that very large system sizes cannot be simulated.
\bibitem{FGOV2000} C. Frontera, J. Goicoechea, J. Ortin, and E. Vives, J. Comput. Phys.
{\bf  160}, 117 (2000).
\bibitem{note3} Note also that for the scaling collapses it has
been possible to use the same values of the {\it a priori} non-universal
constants A and B' in Eqs. (8) and (9).
\end{thebibliography}
\end{document}